\documentclass[aps,showpacs,twocolumn,amsmath,preprintnumbers,nofootinbib,floatfix,secnumarabic]{revtex4}

\usepackage{epsfig,verbatim,hyperref}
\usepackage{amssymb,amsfonts}
\usepackage{graphicx}

\usepackage[caption=false]{subfig}

\usepackage{color}

\def\art{paper }

\def\scn#1#2{\section{#1}\lb{#2}}
\def\sscn#1#2{\subsection{#1}\lb{#2}}
\def\bfl{\begin{flushleft}}
\def\efl{\end{flushleft}}
\def\bfr{\begin{flushright}}
\def\efr{\end{flushright}}
\def\bc{\begin{center}}
\def\ec{\end{center}}
\def\be{\begin{equation}}
\def\ee{\end{equation}}
\def\bse{\begin{subequations}}
\def\ese{\end{subequations}}
\def\ba{\begin{eqnarray}}
\def\ea{\end{eqnarray}}
\def\baa#1{\begin{array}{#1}}
\def\eaa{\end{array}}
\def\bw{\begin{widetext}}
\def\ew{\end{widetext}}
\def\nn{\nonumber }
\def\lb#1{\label{#1}}
\def\bit{\begin{itemize}}
\def\eit{\end{itemize}}
\def\bco{}
\def\bcs{\begin{cases}}
\def\ecs{\end{cases}}

\def\schrod{Schr\"odinger}

\def\lapl{\nabla^2}
\def\laplprm{\nabla^{\prime 2}}

\def\pDer#1#2{\frac{\partial #1}{\partial #2}}

\def\Sin#1#2{\, \text{sin}^{#1}#2}

\def\vena{\vec{\nabla}}

\def\tf{\sigma}
\def\tfc{Q_\tf}
\def\vol{{\cal V}}
\def\psig{\psi_{0} (\vec r)}
\def\eneg{\tilde\omega_{(0)}}
\def\becomes{:=}
\def\angf{\Upsilon}
\def\nc0{\tilde b_0}
\def\enin{{\cal U}_\Psi}

\def\tps{\Delta T_\Psi}

\def\ttps{\Delta\tilde{T}_\Psi}

\def\U{{\rm U}}
\def\H{{\rm H}}

\begin{document}

\preprint{\small Z. Naturforsch. A  \textbf{73}, 619-628 (2018)   
\quad 
[\href{https://doi.org/10.1515/zna-2018-0096}{DOI: 10.1515/zna-2018-0096}]
}

\title{
On the dynamical nature of nonlinear coupling of logarithmic 
quantum wave equation,
Everett-Hirschman entropy and temperature
}

\author{Konstantin G. Zloshchastiev}
\email{http://bit.do/kgz}
\affiliation{Institute of Systems Science, Durban University of Technology, P.O. Box 1334, Durban 4000, South Africa}


\begin{abstract} 
We study the dynamical behavior of nonlinear coupling
in a quantum wave equation of a logarithmic type.
Using statistical mechanical arguments 
for a large class of many-body systems,
this coupling is shown to be related to temperature which is a
thermodynamic
conjugate to the Everett-Hirschman's quantum information entropy.
A combined quantum-mechanical and field-theoretical model is proposed, which
leads to a logarithmic equation with variable nonlinear coupling.
We study its properties and present arguments regarding its nature
and interpretation, including the connection to Landauer's principle.
We also demonstrate that our model is able 
to describe linear quantum-mechanical systems with shape-changing external potentials.
\end{abstract}

\date{
30 Oct 2017 [AdP], 
26 Feb 2018 [ZNA]}


\maketitle

\section{Introduction}

Assuming three spatial dimensions
(lower-dimensional cases can be studied by analogy),  
let us consider the 
logarithmic quantum wave equation, 
often referred also as
the logarithmic Schr\"odinger equation (LogSE):
\ba
i\hbar \partial_t \Psi
&=&
\left(
\hat{{\bf H}} 
-b 
 \ln(a^{3} |\Psi|^{2}
\right) \Psi
\nn\\&=&
\left[-\frac{\hbar^{2}}{2 m} \lapl
+V_\text{ext}
-b 
 \ln(a^{3} |\Psi|^{2})
\right]\Psi
,\label{e:o}
\ea
where 
the coupling 
\be\lb{e:lcoup}
b = b (\vec r, t)
\ee
quantifies the strength of nonlinear self-interaction,
$a$ is a constant parameter of dimensionality length required to make the argument of the logarithm dimensionless,
$m$ is the particle's mass,
$V_\text{ext} = V_\text{ext} (\vec r, t)$ is an external potential (sometimes dubbed
the trap potential),
and
$\lapl = \vec{\nabla} \cdot \vec{\nabla}$ is the 
Laplacian 
(in the relativistic version of Eq. (\ref{e:o}), the derivative part would be replaced by the d'Alembertian).
The wavefunction
$\Psi=\Psi(\vec r, t)$ is 
assumed to be
normalized to 
the number $N$:
\be\lb{e:norm}
\left\langle \Psi | \Psi\right\rangle
\equiv
\int\limits_\vol |\Psi|^2 d^3 \vec r  = N
\geqslant 1
,
\ee
where
$\vol$ is the volume occupied by our system.

The simplest case of Eq. (\ref{e:lcoup}),
\be\lb{e:lcoup0}
b (\vec r, t) = b_0 = \text{const},
\ee
was historically the first to be studied \cite{ros69,ros69a,gg1,gg1a}.
The corresponding models were proven to be  instrumental
in dealing with
extensions of quantum mechanics \cite{gg1,gg1a,bbm79},
physics of quantum fields
and particles \cite{ros69,ros69a,em98,hkt10,Dzhunushaliev:2012zb,gul14,gul15,dmz15}, 
 optics and
transport or diffusion phenomena \cite{gg2,gg6}, 
classical hydrodynamics of Korteweg-type materials \cite{ko1901,ds85,gg5,gg5a,gl08,gl08a}, nuclear
physics \cite{gg3,gg4}, theory of dissipative systems
and quantum information \cite{gg7,gg9,gg10,gg11,lo04,lm13,mw14,zrz17},
theory of quantum liquids and superfluidity \cite{az11,zlo12,bo15,btl16,z17zna}, and theory of
physical vacuum and classical and quantum gravity
\cite{Zloshchastiev:2009zw,gg15,gg14,szm16}.
The mathematical properties of the logarithmic wave equation and its modifications and solutions 
were also
extensively studied  
\cite{bbm79,az11,bo15,z17zna,ch80,hh13,gs13,dms14,ss15,dsz15,js16,ard16,ard16a,ard16b,wct16,tz17,brz17,ns17,fon17,pg17,pa17,zrz17,sha17,as17,bcs18}, to mention just a few examples.

Notwithstanding the success of models with a constant coupling (\ref{e:lcoup0}), 
there remain a few questions which need
to be addressed.

First, what is the value of the coupling (\ref{e:lcoup}), is it a fundamental constant, or is it related to dynamical observables? 
Past studies \cite{gg1,gg1a,gg3,gg4,gkz81} have shown that for a large class of conservative quantum
systems 
the nonlinear term's effect must be rather small.
On the other hand, in the theory of superfluid He-4,
which is an example of a system being 
in thermal contact with a reservoir of large heat capacity as to maintain constant temperature,
this term plays a crucial role \cite{zlo12}.
Therefore, it seems that this coupling can take different values, depending on prevailing
physical conditions,
i.e., it can vary from system to system.
In other words, this means that its value cannot be a fundamental constant, but rather a
dynamical function;
while the formula (\ref{e:lcoup0}) can be regarded as a first-order approximation or 
a limit value of this function.

Second, if this coupling is a nontrivial dynamical notion, then what is its physical meaning?
One possible idea, which was advocated in Refs. \cite{gg9,az11}, is to relate it to 
a certain kind of temperature 
$T_\Psi$,
which 
is a thermodynamical conjugate to a 
quantum analogue of Shannon information entropy \cite{sha48,sha48a}, 
referred here as
the Everett-Hirschman's (EH) entropy \cite{ever55,hir57}.
The latter being defined
as
\be\lb{e:shent}
S_\Psi = -  \langle \Psi | \ln{(a^3 |\Psi|^2)}|\Psi \rangle
=
-\int\limits_\vol |\Psi|^2 \ln{(a^3 |\Psi|^2)} d^3 \vec r
,
\ee
where the Boltzmann constant is hereafter assumed to be $k_B = 1$,
and we adopt the sign conventions of Ref. \cite{sha48,sha48a}.
This entropy can be used as a measure of the localization of a system,
or as an inverse measure of its extendedness, more details can be found
in Ref. \cite{ever55}.

In this framework,
the logarithmic quantum wave equation can be viewed as 
a minimization condition, not for the energy of a system $\langle \Psi | \hat{{\bf H}} | \Psi \rangle$,
but for its quantum
``internal energy'', which is
\be\lb{e:freee}
\enin =
\langle \Psi | \hat{{\bf H}} | \Psi \rangle +  
\tps \, S_\Psi 
, \ \
\ee
where
$T_\Psi$ 
is referred as the \textit{EH} or
\textit{quantum temperature} from now on;
without loss of generality, one can assume
it to be counted with respect
to some reference value.
The standard thermodynamic arguments yield: 
\be\lb{e:tps}
\tps =
\left(\pDer{\enin}{S_\Psi}\right)_\vol
\propto b
,
\ee
where the conventional thermodynamic notations are used. 
In the right-hand side of Eq. \eqref{e:freee},
the term $\langle \Psi | \hat{{\bf H}} | \Psi \rangle$ comes from the system's dynamics, while
the other term
determines the cost of the energy needed
to obtain and handle
information about a system 
\be
I_\Psi = -\log_2 (a^3 |\Psi|^2) = -  \ln (a^3 |\Psi|^2)/ \ln 2
.
\ee
These entropy considerations result in uncertainty relations which can complement the Heisenberg 
relation \cite{ever55,hir57,bab61,beck75}.
The physical meaning of the EH conjugate temperature is discussed in the following sections,
where it is related to conventional (thermal) temperature,
whereas the information-handling cost of energy 
$\tps S_\Psi $
lays quantum-mechanical foundations for the Landauer's principle.

Yet another interpretation of the coupling $b$ comes from the irreversible dynamics
described by the Langevin equation.
According to Ref. \cite{gg7}, in semiclassical approximation this coupling is related to the friction coefficient.
Since the latter is essentially a macroscopic notion, it is unclear whether it has
a well-defined analogue in the quantum picture of reality.
Therefore, in what follows we will focus on an interpretation of the nonlinear coupling 
and logarithmic term in terms of $T_\Psi$ and EH entropy.

Third,
while 
it appears that
the Everett-Hirschman entropy's considerations are an important step towards a
better understanding
of the temperature  $\tps$,
we also must consider what the laws 
governing
its dynamical behavior could be?

These three questions are the main subject of this study.
The \art is organized as follows.
In Sec. \ref{s:fnd}, we enumerate different ways of deriving wave
equations with logarithmic nonlinearity, then we focus on statistical mechanical arguments,
and derive a relation between nonlinear coupling and temperature. 
In Sec. \ref{s:field}, we present a model where nonlinear coupling (hence temperature)
becomes a dynamical value and introduce the basic notions and equations
which we will use in
what follows.
In Sec. \ref{s:prop}, we analytically study some properties and solutions of the 
model.
Discussion and conclusions are presented in Sec. \ref{s:con}, where we also 
outline possible directions
for future research.

\scn{Foundations}{s:fnd}

A wave equation with logarithmic nonlinearity can be introduced into physics using different independent
approaches: dilatation covariance \cite{ros69,ros69a}, 
nonlinear generalization of quantum mechanics preserving energy additivity \cite{gg1,gg1a},
classical Korteweg fluids \cite{gg5,gg5a,gl08,gl08a},
irreversible Langevin dynamics \cite{gg7},
measurement and information entropy \cite{gg9},
effective nonlinearities in quantum systems \cite{zrz17},
superfluidity of helium-4 \cite{zlo12},
theory of physical vacuum, quantum gravity and superfluid-gravity correspondence \cite{Zloshchastiev:2009zw,gg15,szm16}, 
to mention only the examples known to the author.
In this section, we will present another way in which this equation can manifest, one 
underlying some of the above-mentioned approaches.
This new method of deriving the logarithmic nonlinearity will be based on
physically plausible arguments applicable to a large class of many-body systems.

Let us consider a many-body system of particles, whose average potential energy is 
larger than its kinetic (examples 
would be systems made of
strongly interacting particles, or
materials with suppressed kinetic degrees of freedom, such as cold Bose liquids \cite{zlo12}
or melted thermal insulators in capillary tubes \cite{gg5}). 
Then the probability of a microstate is given by a Boltzmann rule,
in which kinetic energy can be neglected
in the leading approximation:
\be\lb{e:statph}
P \propto \exp{(-{\cal E}/T)} \approx \exp{(-\U/T)},
\ee 
where $T$, ${\cal E}$ and $\U$ are, respectively, the temperature, energy and potential energy of a many-body system.

Generally, such a system is described by a large number of linear \schrod~equations;
however, collective degrees of freedom are known to occur in many systems of this kind,
which can substantially simplify the theory \cite{ps04}.
Therefore, if we want to 
effectively describe our system by a single equation,
we must
associate the probability $P$ with 
the wavefunction $\Psi$,
which describes collective degrees of freedom,
and
take into account this statistical effect upon
energy of a system. 
Therefore, we assume 
$|\Psi|^2 \sim P \sim \exp{(-\U/T)}$,
from which a general expression for 
the operator of potential $\U$ 
follows:
$
\hat \U = - K (T-T_0) \ln{(A |\Psi|^2)},
$ 
where $T_0$ is a reference temperature, 
and $A$ and $K$ are some scale constants.
Thus,
in a position representation,
one must include
an additional term,
\be\lb{e:uop}
\langle x | \hat \U |\Psi \rangle
=
\hat \U \Psi (\vec r, t) 
=
- K (T-T_0)
 \ln{(A |\Psi (\vec r, t)|^2)} \Psi (\vec r, t)
,
\ee
into the potential part of the initially linear evolution equation for our system.
If the system is localized inside a vessel or external potential
$V_\text{ext}$, 
then a corresponding term $V_\text{ext} \Psi (\vec r, t)$ must also be added
to a wave equation.

For quantum Hamiltonian systems,
this wave equation can be written in a standard way:
\be\lb{e:wegen}
\hat \H |\Psi \rangle = 
\left(
\frac{\hat{\vec{p}}^{\; 2}}{2 m} 
+ \hat \U + V_\text{ext}
\right) |\Psi \rangle
,
\ee
where $\hat \H = i \hbar \partial_{t}$,
$\hat{\vec{p}} = - i \hbar \vena$ is a momentum operator in a position representation,
and
$m$ is an effective mass of a system's collective degree of freedom.

Finally, 
after redefining proportionality coefficients, 
Eqs. \eqref{e:uop} and \eqref{e:wegen} 
bring us to Eq. (\ref{e:o})
where 
\be\lb{e:ttpsi}
b \sim T \sim T_\Psi
,\ee
where we also recalled the relation 
\eqref{e:tps}.
This formula 
indicates that nonlinear coupling is not a fundamental constant,
but a dynamical value related to physical observables, such as temperature.
Since the latter can generally be a function of position and time, this justifies the necessity 
of studying logarithmic models with a variable $b$, which will be done in subsequent sections.

A final remark can be added here about
the other popular wave equation in the theory of Bose condensates, 
the Gross-Pitaevskii (GP) equation \cite{gp61,gp61a}, sometimes referred as the cubic \schrod~equation, related to the Ginzburg-Landau theory.
This equation, as well as other higher-order polynomial nonlinear \schrod~equations
arising in the theory of condensed Bose systems, is merely one of perturbative limits
of the logarithmic quantum wave equation.
This can be demonstrated by expanding a variational functional corresponding to Eq. \eqref{e:o}
into the Taylor series in the vicinity of its potential's nontrivial extrema, 
see Refs. \cite{gg15,az11} for details.
Indeed, the Taylor series expansion
of a logarithmic term 
in the vicinity of $|\Psi_\text{ext}| = a^{-3/2}$
yields
\be
b 
 \ln\left(a^3 |\Psi|^{2}\right)\Psi
=
b 
\left(
a^3 |\Psi|^2
-
1
\right) \Psi
+ 
\ldots ~
,
\ee
where the leading-order term can be recognized 
as the cubic or Gross-Pitaevskii nonlinearity.
One can also show that the GP equation describes a special case of dilute Bose-Einstein condensates 
where the interparticle interaction potential can be
approximated by a two-body potential
of a contact (delta-singular) shape \cite{ps04}.


\scn{The model}{s:field}

Given that we want to upgrade the nonlinear coupling $b$
(hence temperature, according to the previous section)
to a dynamical value,
we introduce an auxiliary field
$\tf = \tf (\vec r, t)$,
and define the coupling as its scalar function:
$b = b (\tf)$.
Due to the expected gauge invariance of the model, we assume that
the resulting wave equation 
must depend not on the field $\tf$ 
itself, but on its derivative such as the gradient $\vec{\nabla} \tf$.
Since the latter is a vector, whereas the coupling must be a scalar function, we assume
\be\lb{e:bnabla}
b = b \left(\vec n \cdot \vec{\nabla} \tf ,\, \vec{\nabla} \tf \cdot \vec{\nabla} \tf, ... \right)
,
\ee
where $\vec n = \vec r / r = \vec r / \sqrt{\vec r \cdot \vec r}$
is a normal radius vector, and
$r = \sqrt{\vec r \cdot \vec r}$ is an absolute value of the radius vector.

\sscn{Minimal model}{s:mm}

With assumptions \eqref{e:bnabla} 
in hand, we keep only the terms which are linear with respect to $\vec{\nabla} \tf$.
Thus, we introduce the simplest (`minimal')
variable-coupling model as
\ba
i\hbar \partial_t \Psi
&=&
\left[-\frac{\hbar^{2}}{2 m} \lapl
+V_\text{ext}
-
\vec n \cdot \vec{\nabla} \tf\,
 \ln(a^{3} |\Psi|^{2})
\right]\Psi
,~~~~\label{e:vc}
\\
\lapl \tf 
&=&
4 \pi \kappa \rho_\tf,
\lb{e:tf}
\ea
where  
$\rho_\tf = \rho_\tf (\vec r, t)$ is the energy density of the field $\tf$,
and $\kappa$ is a scale constant.
It is natural to assume that this field's distribution is correlated with our system,
therefore, we can impose
\be\lb{e:rhoass}
\kappa\rho_\tf 
= 
f (\rho)
,
\ee
where 
$f$ is a function which must be specified according to a particular model's choice,
and $\rho = |\Psi |^2$ is the probability density of our system;
in the
case of many-body systems ($N \gg 1$), $\rho$ would be an actual particle density.
The exact form of the function $f$ is generally unknown,
and the resulting model (\ref{e:vc})-(\ref{e:rhoass})
is not only nonlinear but also coupled,
therefore
further
analytical studies could become complicated.
Fortunately, some robust simplifications can be made in order
to extract essential physical information.

Henceforth, we focus on the case of a trapless system,
$V_\text{ext} \equiv 0$,
therefore, one can assume a spatial isotropy
of the auxiliary field.
We thus set 
\be
\tf = \tf (r,t),
\ee
hence Eqs. (\ref{e:vc}) and (\ref{e:tf}) can be rewritten
as:
\ba
&&
i\hbar \partial_t \Psi
+\frac{\hbar^{2}}{2 m} \lapl
\Psi
+
\partial_r \tf
 \ln(a^{3} |\Psi|^{2})
\Psi
=
0
,~~\label{e:vc2}
\\
&&
\lapl \tf 
=
\lapl_r \tf 
=
4 \pi 
f(|\Psi |^2)
,
\lb{e:tf2}
\ea
where
$
\lapl_r 
=
\partial^2_{r,r} 
+
\frac{2}{r}
\partial_{r} 
$ is a radial part of Laplacian.
These two equations must be supplemented with 
the normalization condition for $\Psi$, boundary conditions for 
both $\Psi$ and $\tf$, and 
a specific expression for a function $f$ depending
on the physical system in question.

\sscn{Approximate minimal model}{s:mmapp}

The minimal model \eqref{e:vc2}, \eqref{e:tf2}
contains the function $f$ whose value must be specified depending
on the physical system in question, otherwise it is generally unknown.
However, a certain
class of dynamical systems must have a common $f$, at least
in the leading approximation.
 
For simplicity, let us impose here the time independence of the auxiliary field: 
$\tf = \tf (r)$.
Furthermore,
let us assume that 
our system
satisfies the condition:
$\lim\limits_{r\to+\infty}  
\rho_\tf 
= \lim\limits_{r\to+\infty}  |\Psi |^2
= 0$, 
therefore
its density
can be formally represented
as a decomposition of the Dirac delta-singular
part
and asymptotically vanishing extended part:
\be
\rho_\tf 
\propto
f(|\Psi |^2)
\sim
\delta (r)
+
\sum\limits_{n = 1}^\infty
\frac{a_n}{r^n}
,
\ee 
where 
$\delta (r)$ is the Dirac's delta function centered in the origin.
This expression can be viewed as describing a point-like object's density, 
plus an extended object's density distribution represented by a Taylor series expansion
with respect to $1/r$, which ensures that the field $\tf$ vanishes at spatial infinity.

Under these assumptions,
Eq. (\ref{e:tf2}) can be approximately written in series form
and decoupled from Eq. (\ref{e:vc2}):
\be
\lapl_r \tf 
=
-4 \pi 
\left[
\tfc \delta (r)
-
\frac{b_0}{2 \pi r}
+
{\cal O} (1/r^2)
\right]
,
\lb{e:tfapp}
\ee
where
$\tfc = -
\tfrac{1}{4 \pi 
} 
\oint \vec\nabla\tf \cdot d \vec S$ and $b_0$ are constants (the former being a Gauss law's charge),
and notation
${\cal O} (1/r^2)$ represents terms which decay faster than $1/r$ when $r \to \infty$.
Here, the constant $\tfc$ labels the delta-singular part of the density $\rho$,
whereas the series coefficient $b_0$ labels the leading-order term of the extended part;
the chosen notation $b_0$ is not a coincidence, as we will see below.

Notice that in the original model (\ref{e:vc2})-(\ref{e:tf2}), both $\tfc$ and $b_0$
would not be built-in parameters of a theory, 
but integration constants, therefore their values would depend on boundary conditions,
and could therefore vary from system to system.
Thus, the full model would allow us to reduce the number of parameters of a logarithmically nonlinear theory and make
it more self-contained.
However, within the frameworks of the approximation (\ref{e:tfapp}), 
values of $\tfc$ and $b_0$ are unknown, and have yet to be determined from other factors. 

Furthermore,
neglecting higher-order terms ${\cal O} (1/r^2)$,
we can exactly solve Eq. (\ref{e:tfapp}).
We obtain
\be
\tf = \tf_0 + \frac{\tfc}{r} + b_0 r
,
\ee
where $\tf_0$ is an additive constant, which can be set to zero due to the gauge invariance
of the $\tf$-field.
Substituting this into Eq. (\ref{e:vc2}),
we obtain,
\be
i\hbar \partial_t \Psi
+
\frac{\hbar^{2}}{2 m} \lapl
\Psi
+
\left(
b_0 - \frac{\tfc}{r^2}
\right)\!
 \ln(a^{3} |\Psi|^{2})
\Psi
=0
,\label{e:vcm}
\ee
thus confirming our earlier expectations that the nonlinear coupling is not generally constant.
If the $\tf$ field's charge $\tfc$ is nonzero 
then at  $r \to 0$ the coupling's magnitude
grows like $1/r^2$, whereas at large $r \to \infty$ the coupling tends to a constant, so that 
one asymptotically recovers Eq. (\ref{e:lcoup0}).

Notice that the constant part of the coupling, $b_0$,
is induced not by the delta-singular part of the
$\tf$-field's distribution but by its extended part.
This explains why solutions of the conventional logarithmic equation
(\ref{e:lcoup0}) are applicable for describing non-singular extended objects,
such as Q-balls and finite-size particles \cite{em98,hkt10,Dzhunushaliev:2012zb,gul14,gul15,dmz15}
and superfluid droplets \cite{az11,zlo12}.
Additionally, the appearance of a new term, proportional to $\tfc$, indicates
that the new model 
could also be instrumental in dealing with singular or point-like objects.

For the calculations that follow, we will make Eq. (\ref{e:vcm}) dimensionless.
This equation always contains three constants, $\hbar$, $m$ and $a$,
which are independent of the nonlinear coupling $b$.
Therefore, from them one can construct the following scales of length, time and mass, respectively:
$
a
$, $
m a^2/\hbar
$, $
m
$.
Assuming $a >0$ and
\be
\vec r \,'= \vec r/a,\
t' = t/\tau
,\ 
\tilde\Psi = a^{3/2} \Psi
,
\ee
where
$
\tau = 2 m a^2/\hbar
$,
we can write Eq. (\ref{e:vcm}) in a dimensionless form:
\be
i \partial_{t'} \tilde\Psi
+
\laplprm \tilde\Psi
+
\left(
\nc0 - \frac{\tilde q}{r^{\prime 2}}
\right)\!
 \ln(|\tilde\Psi|^{2})
\tilde\Psi
=
0
,\label{e:vcmd}
\ee
where 
$\nc0 = b_0 \tau/\hbar = 2 m b_0 a^2 / \hbar^2$ 
and 
$\tilde q = \tfc \tau/ (\hbar a^2) = 2 m \tfc / \hbar^2 $.
In the following sections we will omit primes, assuming
that times, lengths, momenta and energies are measured in units
of $\tau$, $a$, $\hbar/a$ and $\hbar/\tau$, respectively.

\scn{Properties and solutions}{s:prop}

In this section,
we consider a stationary case
and analytically derive corresponding solutions. 
We begin by imposing
a stationary 
ansatz
\be\lb{e:sta}
\tilde\Psi (\vec r, t) = \exp{(- i \tilde\omega t)} \psi (\vec r),
\ee
where $\tilde \omega$ is 
a frequency measured in units of $1/\tau$.
Then Eq. (\ref{e:vcmd}) becomes
an eigenvalue equation for this frequency: 
\be
\lapl
\psi
+
\left(
\nc0 - \frac{\tilde q}{r^{2}}
\right)\!
 \ln(|\psi|^{2})
\psi
+
\tilde\omega \psi
=
0
,\label{e:vcmds}
\ee
where $\psi = \psi (\vec r)$ is a spatial wavefunction
normalized to a number $N$: 
$\int_\vol |\psi|^2 d^3 \vec r  = N \geqslant 1$.

Due to the symmetry of Eq. (\ref{e:vcmds}), it is
convenient to work in spherical coordinates from now on.
Then the Laplacian can be decomposed into its radial and angular parts
\ba
\lapl = \lapl_r + 
\frac{1}{r^2} \lapl_{S^2}
,\lb{e:lapdcmp}
\ea
where 
$
\lapl_{S^2}
=
\tfrac{1}{\sin\theta}
\partial_\theta 
\left(
\sin\theta \, \partial_\theta
\right)
+
\tfrac{1}{\Sin{2}{\theta}}
\partial^2_{\varphi,\varphi}
$
is the Laplace-Beltrami operator on a sphere.

Here we also introduce the notion of effective external potential.
Once a solution of Eq. (\ref{e:vcmds}) is known,
the effective external potential for such a solution 
is given, in a dimensionless form,
by the expression:
\be\lb{e:veff}
\tilde V_\text{eff} (\vec r)
=
\left(
\frac{\tilde q}{r^{2}}
-
\nc0
\right)\!
 \ln(|\psi_s (\vec r)|^{2}) 
,
\ee
where $\psi_s (\vec r)$ is a solution's wavefunction.
This potential indicates that a solution $\psi_s$ can be equivalently
derived 
from the linear \schrod~equation
with 
external potential $V_\text{ext} = V_\text{eff} (\vec r)$.
In other words, an observer would not be able to empirically differentiate
a nonlinear problem from a linear one, if based on the analysis
of a solution $\psi_s$ alone.
Note that the effective potential's shape would vary from solution to solution
for the same system, therefore, a nonlinear theory of type (\ref{e:vc2}), (\ref{e:tf2})
has the capacity 
to describe linear systems with shape-changing external potentials
depending on a state, e.g.,
those systems which undergo phase transitions as their temperature changes.
Besides,
this creates a framework for creating quantum-mechanical models where an 
external potential
is not \textit{ab initio} postulated but actually derived. 

Another instrumental value to be introduced is a radial density
of the Everett-Hirschman entropy \eqref{e:shent},
measured in units $1/a$:
\be\lb{e:shentrad}
\tilde s_\Psi^\text{(r)} 
=
-\iint |\psi|^2 \ln{(|\psi|^2)} r^2 \Sin{2}{\theta} d \theta d \phi
,
\ee
where the integral is taken over the sphere;
then the entropy (\ref{e:shent})
is simply 
$S_\Psi = \int_0^{\infty} s_\Psi^\text{(r)} d r$.
The $S_\Psi$-conjugate temperature in this case is:
\be\lb{e:tss}
\ttps
=
\nc0
-
\frac{\tilde q}{r^{2}} 
,
\ee
when written in our units of energy $\hbar/\tau$. 

Furthermore, when dealing with analytical solutions of Eq. (\ref{e:vcmds}),
one must distinguish between different cases of 
nonlinear couplings' parameters that occur:

\sscn{Case $\nc0 \not= 0$, 
$\tilde q 
\not= 0, 1
$
}{s:c1}

In this case, 
the normalized spherically-symmetric solution of Eq. (\ref{e:vcmds}) 
can be written as:
\be\lb{e:psicgen}
\psig =
\exp{\!\left(
- \frac{\pi}{2 N^{2/3}}r^2 \right)}
,
\ee
and
both $\tilde\omega$ and coupling $\nc0$ are no longer arbitrary, 
but
become eigenvalues:
\ba
\tilde\omega \becomes
\eneg
&=&
\nc0
(3 - \tilde q)
=
\frac{\pi (3 - \tilde q)}{N^{2/3}}
,\lb{e:omcgen}\\
\nc0 \becomes
\tilde b_{0 (0)}
&=&
\frac{\pi}{N^{2/3}}
,\lb{e:bcgen}
\ea
where the subscript `(0)' denotes the ground state.

Equations \eqref{e:omcgen} and \eqref{e:bcgen}
indicate that for this solution to exist, the original parameters $m$, $b_0$
and $a$ must not be independent, but must obey a constraint
$2 m b_0 a^2  = \pi \hbar^2/N^{2/3}$ instead.
For the model's applications, this can be helpful because it decreases the number of free parameters.
Notice also that the (eigen)value
of 
$b_{0}$ depends on the combination of other parameters, namely
$m a^2 N^{2/3}$, which could explain empirical non-observability of logarithmic nonlinear
effects in some systems and their dominance in others; further discussion of this can be found
in the concluding section.

For the solution (\ref{e:psicgen})-(\ref{e:bcgen}),
the effective external potential (\ref{e:veff}) can be evaluated as
\be\lb{e:veffcgen}
\tilde V^{(0)}_\text{eff} (\vec r)
=
\frac{1}{4}
\tilde\Omega_\text{eff}^2
\left(r^2 - \tilde q_N \right)
,
\ee
where 
$\tilde\Omega_\text{eff} = 2 \tilde b_{0 (0)} = 2\pi/N^{2/3}$
and
$\tilde q_N = \tilde q / \tilde b_{0 (0)} = \tilde q N^{2/3}/\pi$.
This formula indicates that 
most of physical properties of the solution (\ref{e:psicgen})-(\ref{e:bcgen})
must be identical to those
of a quantum harmonic oscillator of the 
dimensionless frequency $\tilde\Omega_\text{eff}$.
This correspondence between LogSE's ground states and quantum harmonic oscillators was noticed in Ref. \cite{zrz17}.
However, excited states are not likely to be interpreted in terms of the
oscillator (\ref{e:veffcgen}),
because the corresponding expression for an effective external potential would certainly be more complex.

Furthermore,
the entropy density (\ref{e:shentrad}) for the solution (\ref{e:psicgen})-(\ref{e:bcgen}) appears to be
\be
\tilde s_\Psi^\text{(r)}
=
\frac{4 \pi^2}{N^{2/3}} 
r^4
\exp{\!\left(
- \frac{\pi}{N^{2/3}}r^2 \right)}
,
\ee
and the integral Everett-Hirschman entropy \eqref{e:shent} is simply
\be
S_\Psi
=
\frac{3}{2} N
,
\ee
which results in the following relation between $S_\Psi$ 
and frequency's eigenvalue for this solution
which does not contain the normalization number:
$
\eneg
S_\Psi^{2/3}
=
\pi
\left(
3/2
\right)^{2/3}
(3 - \tilde q)
$,
where the frequency $\eneg$ is defined in Eq. (\ref{e:omcgen}).

The quantum temperature (\ref{e:tss}) 
becomes in this case:
\be
\ttps
=
\frac{\pi}{N^{2/3}}
\left(
1
- \frac{\tilde q_N}{r^{2}}
\right)
.
\ee
The value $\ttps$ is always positive-definite if $ \tilde q \leqslant 0$
which corresponds to a non-negative value of the charge $\tfc$.
If $ \tilde q > 0$ then the sign of $\ttps$ changes when
crossing the radius $\sqrt{\tilde q_N}$ and becomes negative at
$r < \sqrt{\tilde q_N}$.
Possible reasons for, and implications of this behavior are discussed in the
concluding section.

\sscn{Case $\nc0 \not= 0$, $\tilde q = 1$}{s:c2}

In this case, 
the normalized solution of Eq. (\ref{e:vcmds}) 
and a corresponding frequency eigenvalue
can be written as, respectively:
\ba
\psig 
&=&
\exp{\!\left(
\tilde k r 
- \frac{1}{2} \nc0 \, r^2 \right)}
,\lb{e:psicq1}
\\
\tilde\omega \becomes
\eneg
&=&
2 \nc0 - \tilde k^2
,
\ea
where the constant $\tilde k$ 
is a solution of the transcendental equation
\be
\sqrt{\frac{\pi}{{\nc0}}}
\left(
\frac{\nc0}{2} +  \tilde k^2
\right)
\left[
1 + \text{erf}\!\left(\frac{\tilde k}{\sqrt{\nc0}}\right)
\right]
\text{e}^{\tilde k^2/\nc0}
=
\frac{N \nc0^{2}}{2\pi}
- \tilde k
,
\ee
while $\nc0$ remains a free parameter.

For the solution (\ref{e:psicq1}),
the effective external potential (\ref{e:veff})
becomes
\be\lb{e:veffcq1}
\tilde V^{(0)}_\text{eff} (\vec r)
=
\frac{2 \tilde k}{r}
+
\nc0^2 
\left(
r -  \frac{\tilde k}{\nc0}
\right)^2
- \nc0  -  \tilde k^2
,
\ee
thus indicating that 
most of physical properties of the system in a ground state 
would be identical to those
of a particle 
trapped
in the harmonic potential of a frequency $2 \nc0$
crossed with the Coulomb-type potential with a strength constant $2 \tilde k$.
Notice also the change of the effective potential's shape compared
to the previous case (\ref{e:veffcgen}) where it is purely harmonic.

The entropy density (\ref{e:shentrad}) for the solution (\ref{e:psicq1}) can be
evaluated as
\be
\tilde s_\Psi^\text{(r)}
=
4 \pi 
r^3
(\nc0 r - 2 \tilde k)
\exp{\!\left(
2 \tilde k r - \nc0 r^2
\right)}
,
\ee
and the  Everett-Hirschman entropy is 
\be
S_\Psi
=
\frac{
N (3 \nc0^2 - 4 \tilde k^4) - 4 \pi \tilde k
}{
2 \nc0 (\nc0 + 2 \tilde k^2)
} 
.
\ee
The quantum temperature (\ref{e:tss}) 
becomes in this case:
\be
\ttps
=
\nc0
-
\frac{1}{r^{2}}
,
\ee
thus,
similarly to the previous case,
$\ttps$ is positive at large $r$,
changing sign when crossing the radius $\sqrt{\nc0}$,
and becomes negative at small $r$. 

\sscn{Case $\nc0 \not= 0$, $\tilde q = 0$}{s:c3}

In this case, we recover the logarithmic \schrod~equation with a constant
nonlinear coupling, 
\be
i \partial_{t} \psi
+
\lapl
\psi
+
\nc0 
 \ln(|\psi|^{2})
\psi
=
0
,\label{e:logcon}
\ee
discussed after Eq. (\ref{e:lcoup0}) above.

For a stationary case (\ref{e:sta}), the normalized spherically-symmetric ground-state solution 
of Eq. (\ref{e:logcon}) 
and a corresponding frequency eigenvalue
can be written as, respectively:
\ba
\psig 
&=&
\left(\frac{\nc0}{\pi}\right)^{3/4}
\sqrt N
\exp{\!\left(
- \frac{1}{2} \nc0\, r^2 \right)}
,
\lb{e:psiccon}\\
\tilde\omega \becomes
\eneg
&=&
3 \nc0 
\left[
1 - 
\frac{1}{2}
\ln{\!\left(\frac{\nc0 N^{2/3}}{\pi}\right)}
\right]
, \lb{e:omccoup}
\ea
which describes a Gaussian-shaped spherical wave.

For the solution (\ref{e:psiccon}),
the effective external potential (\ref{e:veff}) reads
\be\lb{e:veffccon}
\tilde V^{(0)}_\text{eff} (\vec r)
=
\nc0^2 r^2
,
\ee
which makes this case similar to Eq. (\ref{e:veffcgen}):
most of physical properties of the solution (\ref{e:psiccon})
must be determined by a quantum harmonic oscillator of the mass $1/2$ 
and dimensionless frequency $2 \nc0$.

The entropy density (\ref{e:shentrad}) for this solution appears to be
\be
\tilde s_\Psi^\text{(r)}
=
\frac{4 \nc0^{3/2} N}{\sqrt\pi}
r^2
\left[
\nc0 r^2
-
\frac{3}{2} \ln{\!\left(\frac{\nc0 N^{2/3}}{\pi}\right)}
\right]
\text{e}^{
- \nc0 r^2}
,
\ee
hence the integrated Everett-Hirschman entropy is simply
\be
S_\Psi
=
\frac{3}{2} 
\left[
N
-
\ln{\!\left(\frac{\nc0 N^{2/3}}{\pi}\right)}
\right]
=
N 
\left(
\frac{9}{2}
- \frac{\eneg}{\nc0}
\right)
,
\ee
where the frequency $\eneg$ is defined in Eq. (\ref{e:omccoup}).
This results in the following relation between $S_\Psi$ 
and frequency's eigenvalue for this solution:
\ba
\eneg
-
\frac{
\pi (9 N - 2 S_\Psi)
}{
2 N^{5/3}
} 
\exp{
\left(
\frac{2 S_\Psi}{3 N} -1
\right)
}
=0
.
\ea
The conjugate quantum temperature (\ref{e:tss}) 
becomes a constant in this case 
\be
\ttps
=
\nc0
,
\ee
which is positive-definite.

\sscn{Case $\nc0 = 0$, $\tilde q \not = 0$}{s:csep}

In this case, Eq. (\ref{e:vcmd}) becomes
\be
i \partial_{t} \psi
+
\lapl \psi
- \frac{\tilde q}{r^{2}}
 \ln(|\psi|^{2})
\psi
=
0
,
\label{e:logsep}
\ee
where a decomposition \eqref{e:lapdcmp} is implied.

While this equation looks more complicated to solve
than the logarithmic equation 
with a constant nonlinear coupling (\ref{e:logcon}),
it has certain features which make it easier to study.

Most significantly, this equation allows a separation of angular variables
from others, which paves the way for us to drastically decrease the
dimensionality of the problem in a general case.
Assuming the stationary ansatz (\ref{e:sta}),
where
\be
\psi (\vec r)
=
R (r) \, \angf (\theta,\varphi)
,
\ee
and
using the decomposition (\ref{e:lapdcmp}),
we can separate Eq. (\ref{e:logsep})
into its radial and angular parts:
\ba &&
\lapl_r R
-
\frac{1}{r^2} 
\left[
L^2
+ \tilde q
 \ln(|R|^{2})
\right]
R
+
\tilde\omega R
=
0
,\label{e:logseprad}
\\&& 
\lapl_{S^2} \angf 
- \tilde q
 \ln(|\angf|^{2}) \angf
+ L^2\, \angf = 0
,
\label{e:logsepang}
\ea
where
$L$ is a separation constant whose eigenvalue follows from the last equation.
The latter resembles a differential equation for spherical harmonics, 
but
contains
a nonlinear term, making its solutions a separate topic of research.
This equation indicates that 
the system's total angular momentum 
acquires a nonlinear correction which can manifest  
in those systems for which $\tfc \not= 0$.

Furthermore,
equation (\ref{e:logsep}) can be viewed either as a short-distance
limit 
$r \ll \sqrt{\tilde q/\nc0} = \tfc/(b_0 a^2)$
of Eq. (\ref{e:vcmd}),
or as a large-charge limit, $\tfc \to \infty$, thereof.
It is thus no longer necessary to assume that $b_0$ is unnaturally small 
to fit existing experimental data for physical systems for which the
model (\ref{e:lcoup0}) is \textit{a priori} inapplicable.
It is sufficient to assume that models with a large value of $\tfc$
are more relevant for those systems.

Because the separation of angular variables from others is possible in this case,
we need not restrict ourselves to a spherically symmetric ansatz 
to find analytical solutions.
Instead, we will
search for a solution for the radial wavefunction $R(r)$ given by Eq. (\ref{e:logseprad})
and normalized as $\int_0^\infty |R|^2 r^2 d r = N$.
Then 
the solution of Eq. (\ref{e:logseprad}) 
can be written as:
\ba
R_{0,L} (r)
&=&
\left\{
\baa{rl}
\exp{\!\left(
- \tilde\mu^2 r
- L^2/2 
\right)} 
&
\text{if} \
\tilde q =1 
,\\
\exp{\!\left(
- L^2/2 
\right)}
& 
\text{if} \
\tilde q \not=1
,
\eaa
\right.
\lb{e:psicsep}
\\
\tilde\omega \becomes
\eneg
&=&
\left\{
\baa{rl}
- \tilde\mu^4
&
\text{if} \
\tilde q =1 
,\\
0
& 
\text{if} \
\tilde q \not=1
,
\eaa
\right.
\ea
where 
$
\tilde\mu^2 = 
(4 N)^{-1/3} \exp{(-L^2/3)} 
$,
and
a subscript `0' refers to a lowest value of a main quantum number.
It is clear that on a real semi-axis a physically suitable 
solution is the one for which 
\be\lb{e:tfccsep}
\tilde q = 
1,
\ee
which can be regarded as a constraint for the
original value $\tfc$: $\tfc \to \hbar^2/(2 m)$. 

For the solution (\ref{e:psicsep})-(\ref{e:tfccsep}),
the effective external potential (\ref{e:veff})
reads:
\be\lb{e:veffcsep}
\tilde V^{(0)}_\text{eff} (\vec r)
=
- \frac{2 \tilde\mu^2}{r}
-
\frac{L^2}{r^2}
,
\ee
thus indicating that 
most of physical properties of this solution
would be identical to those
of a particle 
in the 
Coulomb-type potential with a strength constant $2 \tilde\mu^2$,
in presence of 
the standard ``centrifugal'' potential $L^2/r^2$.

Furthermore,
the entropy density (\ref{e:shentrad}) for the solution (\ref{e:psicsep})-(\ref{e:tfccsep})
appears to be
\be
\tilde s_\Psi^\text{(r)}
=
8
\tilde\mu^8 N r^3
\left(
1
+ \frac{L^2 + S_\angf}{2 \tilde\mu^2 r}
\right)
\exp{\!\left(-2 \tilde\mu^2 r\right)}
,
\ee
where we denoted the constant
\[
S_\angf = -
\iint |\angf|^2 \ln{(|\angf|^2)} \Sin{2}{\theta} \, d \theta \, d \phi
,
\]
an integration being taken over a sphere.
The integral EH entropy \eqref{e:shent} reads: 
\be
S_\Psi
=
N (L^2 + S_\angf+ 3)
,
\ee
while the conjugate quantum temperature (\ref{e:tss}) 
becomes in this case simply:
\be
\ttps
=
-
\frac{1}{r^{2}}
,
\ee
implications of which are discussed below.

\scn{Discussion}{s:con}

We have studied the dynamical behavior of nonlinear coupling $b$
in the quantum wave equation of a logarithmic type.
Using statistical mechanics arguments,
we have shown that this coupling is related to the thermal temperature of many-body systems
which
satisfy the following conditions: (i) their potential energy must be 
larger than kinetic, (ii) they must allow an effective description in terms
of collective degrees of freedom governed by a wavefunction. 
One example of such systems would be strongly interacting Bose liquids,
where the logarithmic model was shown to be very instrumental \cite{az11,zlo12}.

Furthermore, the nonlinear coupling has been 
associated with a certain kind of quantum temperature:
a thermodynamical conjugate to the Everett-Hirschman entropy,
which allows us to relate thermal temperature, EH quantum temperature, and nonlinear coupling,
as discussed in Sec. \ref{s:fnd}.

In view of the dynamical nature of nonlinear coupling, a combined quantum-mechanical and field-theoretical model is proposed in Sec. \ref{s:field}, which
leads to a logarithmic equation with variable nonlinear coupling.
By studying the behavior of the latter, one achieves deeper understanding 
of the Everett-Hirschman entropy and its thermodynamical conjugate.
Considering this model in a next-to-leading order approximation with respect
to the auxiliary field $\tf$, see Sec. \ref{s:mmapp}, we analytically obtained a number of stationary solutions
and established a number of profound properties, given in Sec. \ref{s:prop}.
Below we present discussion of the results obtained.

First, 
the logarithmic model reveals a connection between the EH conjugate temperature and
the conventional (thermal) one.
The latter is still a notion which is well-defined at a classical level only
(\textit{e.g.}, as measured 
by bringing a system into a thermal equilibrium with a calibrated system),
but its quantum analogue 
is unknown.
This problem manifests itself most strikingly when dealing with cold quantum gases and liquids such as
Bose-Einstein condensates. 
Strictly speaking, one can not measure the temperature of a condensate 
in a classical way -- experimental condensates are energetically isolated,
so no thermal equilibrium can be achieved with a calibrated system without affecting the condensate's state.
Thus, the standard method of measurement consists 
of switching a trap off,
using a laser upon the condensate's atoms and measuring the scattered light
to deduce their temperature 
from experimental profiles of density and momentum distributions, assuming that the energy spectrum is also known.
However, this method of measurement presumes that relations between temperature and the above-mentioned distributions and spectrum
are derived from some theoretical model, which must be thus presumed to be \textit{a priori} valid
for that particular condensate.
Besides, 
the detrapping measurement  is unlikely to be reliable for 
strongly interacting quantum Bose liquids, which tend
to confine their atoms into a droplet, in absence of trapping potentials \cite{az11,zlo12,z17zna}.

In this regard, a conjecture that thermal temperature is related to quantum temperature $T_\Psi$ and nonlinear 
coupling $b$, at least for a large class of systems,
has been discussed in Sec. \ref{s:fnd}.
Aside from solving the above-mentioned issue with the fundamental (quantum-mechanical) definition of the thermal temperature, 
the relation \eqref{e:ttpsi} 
lays quantum-mechanical foundations also for the Landauer's principle \cite{lan61}.
Namely, it is the term $\tps\, S_\Psi$
from Eq. (\ref{e:freee}) which is responsible for
information-handling cost of energy, including the energy cost for information erase.
The latter has been confirmed in experiments with
different nanoscale systems \cite{bap12,jgb14,hld16}.

Second, 
in the model with variable nonlinear coupling,
the nature of quantum EH temperature (\ref{e:freee}) 
becomes clearer.
For the solution described in Sec. \ref{s:c3}, the value $\tps$ is positive semi-definite,
whereas in Sec. \ref{s:csep} it is negative semi-definite;
in Secs. \ref{s:c1} and \ref{s:c2} it changes its sign at a certain value of radius.
The common feature of the cases in Secs. \ref{s:c1} and \ref{s:c2}
is that $\tps$ tends toward negative values
at small distances from the origin and to a positive constant at large distances.
Moreover,
the common feature of all cases is that $\tps$ tends to negative values
whenever the term $\tfc/r^2$ in the nonlinear coupling predominates over the constant one. 
Analyzing these features together, one can 
hypothesize that the EH temperature can
serve as a means of differentiating phases, \textit{e.g.}, those related to the
microscopic and macroscopic scales of radius:
$\tps$ is negative for the microscopic scale and it is positive for the macroscopic one (up to a sign convention adopted
in the definition of $S_\Psi$).

Third,
in the minimal model of Sec. \ref{s:field},
the coupling constant $b_0$
is no longer a predefined parameter of a theory, cf. Eq. (\ref{e:lcoup0}).
Instead, it becomes
one of the integration constants of evolution equations, such as Eqs. (\ref{e:vc2}) and (\ref{e:tf2}),
therefore, its value can vary from system to system.
Thus, the full model allows us to reduce a number of parameters of the theory and make
it more self-consistent and self-sufficient.
However, within the frameworks of the approximation (\ref{e:tfapp})-(\ref{e:vcm}), 
values of $\tfc$ and $b_0$ are unknown and have yet to be determined from other considerations. 

Nevertheless, even the approximate minimal model, cf. Sec. \ref{s:mmapp}, offers an explanation
as to why the constant $b_0$ is negligible for some systems but crucial
for others, as mentioned in the Introduction.
To illustrate this,
let us compare cases described in Sec. \ref{s:prop}.
For the solutions of Eq. (\ref{e:vcm}), which are described
in Secs. \ref{s:c2} and \ref{s:c3}, the constant $b_0$
remains a free parameter, which can take any value, either defined \textit{ad hoc} or fitted from an 
experiment;
an example of the latter procedure can be found in Ref. \cite{zlo12}.
However, for the solution in Sec. \ref{s:c1}, the constant $b_0$ becomes an eigenvalue, \textit{i.e.}, a function
of other constants of the model and quantum numbers (if one considers excited states).
Specifically, it is small if the combination $m a^2 N^{2/3}$ is large.
Because neither of those three constants are fundamental nor universal for all quantum systems, the value of $b_0$ can vary between systems.
Moreover, it can also vary for different solutions of the same system,
because of the above-mentioned eigenvalue structure and a Hilbert space associated with it.

Finally, effective external potentials computed in Secs. \ref{s:c2} and \ref{s:csep} 
illustrate a possibility
that some fundamental interactions, such as gravity, could emerge  
as a nonlinear quantum-mechanical phenomenon based on a concept of 
the quantum information entropy, cf. Eq.  (\ref{e:freee}),
and evolution equations of a logarithmic type.
This conjecture is supported by other studies,
which suggest that
the most probable candidate for such a phenomenon
is a background superfluid of a logarithmic type \cite{Zloshchastiev:2009zw,gg15,szm16}.

\begin{acknowledgments}
Discussions with I. Sinaysky from University of Kwazulu Natal‎
(who brought the Landauer's work into my attention)
and A. Avdeenkov from I.I. Leypunsky Institute of Physics and Power Engineering
are acknowledged.
Proofreading of the
manuscript by P. Stannard is greatly appreciated.
This work is based on the research supported by the National Research Foundation of South Africa 
under Grants Nos.
95965 
and 98892.

\end{acknowledgments}



\begin{thebibliography}{0}


\bibitem{ros69}
G. Rosen,
J. Math. Phys. \textbf{9}, 996 (1968).

\bibitem{ros69a}
G. Rosen,
Phys. Rev. \textbf{183}, 1186 (1969).

\bibitem{gg1}
I.~Bialynicki-Birula and J.~Mycielski,
Ann. Phys. (N. Y.) {\bf 100}, 62 (1976).

\bibitem{gg1a}
I.~Bialynicki-Birula and J.~Mycielski,
Commun. Math. Phys. \textbf{44}, 129 (1975).


\bibitem{bbm79}
I.~Bialynicki-Birula and J.~Mycielski,
Phys.\ Scripta {\bf 20}, 539 (1979).
	
\bibitem{em98}
K. Enqvist and J. McDonald,
Phys. Lett. B \textbf{425}, 309-321 (1998).
	
\bibitem{hkt10}
T. Hiramatsu, M. Kawasaki, and F. Takahashi, 
J. Cosmol. Astropart. Phys. \textbf{2010}, 008 (2010).

\bibitem{Dzhunushaliev:2012zb}
V.~Dzhunushaliev and K.~G.~Zloshchastiev,
Central Eur. J. Phys. \textbf{11}, 325-335 (2013) [arXiv:1204.6380].

\bibitem{gul14}
I. E. Gulamov, E. Ya. Nugaev, and M. N. Smolyakov,
Phys. Rev. D \textbf{89}, 085006 (2014).

\bibitem{gul15}
I. E. Gulamov, E. Ya. Nugaev, A. G. Panin, and M. N. Smolyakov,
Phys. Rev. D \textbf{92}, 045011 (2015).

\bibitem{dmz15}
V.~Dzhunushaliev, A. Makhmudov, and K.~G. Zloshchastiev,
Phys. Rev. D \textbf{94}, 096012 (2016).




\bibitem{gg2}
H. Buljan, A. \v Siber, M. Solja\v ci\'c, T. Schwartz, M. Segev, and D. N. Christodoulides,
Phys. Rev. E \textbf{68}, 036607 (2003).

\bibitem{gg6}
T. Hansson, D. Anderson, and M. Lisak,
Phys. Rev. A \textbf{80}, 033819 (2009).

\bibitem{ko1901}
D. Korteweg, 
Arch. Neerl. Sci. Exactes Nat. \textbf{6}, 1-24 (1901).

\bibitem{ds85}
J.E. Dunn and J.B. Serrin, 
Arch. Rat. Mech. Anal. \textbf{88}, 95-133 (1985).

\bibitem{gg5}
S. De Martino, M. Falanga, C. Godano and G. Lauro,
Europhys. Lett. \textbf{63}, 472 (2003).

\bibitem{gg5a}
S. De Martino and G. Lauro,
in: Proceed. 12th Conference on WASCOM, 148 (2003).

\bibitem{gl08}
G. Lauro,
Geophys. Astrophys. Fluid Dyn. \textbf{102}, 373-380 (2008).

\bibitem{gl08a}
G. Lauro,
Acta Appl. Math. \textbf{132}, 405 (2014). 


\bibitem{gg3}
E. F. Hefter,
Phys. Rev. A \textbf{32}, 1201 (1985).

\bibitem{gg4}
V.~G.~Kartavenko, K.~A.~Gridnev and W.~Greiner,
  Int.\ J.\ Mod.\ Phys.\  E {\bf 7}, 287 (1998).
	




\bibitem{gg7}
K.~Yasue,
Ann. Phys. (N.Y.)\  {\bf 114}, 479 (1978).


\bibitem{gg9}
J. D. Brasher,
Int. J. Theor. Phys. \textbf{30}, 979 (1991).




\bibitem{gg10}
D. Schuch,
Phys. Rev. A \textbf{55}, 935 (1997).

\bibitem{gg11}
M. P. Davidson,
Nuov. Cim. B \textbf{116}, 1291 (2001).

\bibitem{lo04}
J. L. L\'opez,
Phys. Rev. E \textbf{69}, 026110 (2004).

\bibitem{lm13}
J. L. L\'opez and J. Montejo-G\'amez, 
Nanoscale Syst. Math. Model. Theory Appl. \textbf{2}, 49-80 (2013). 


\bibitem{mw14}
D. A. Meyer and T. G. Wong,
Phys. Rev. A \textbf{89}, 012312 (2014).

\bibitem{zrz17}
M. Znojil, F. R\r{u}\v{z}i\v{c}ka, and K.~G.~Zloshchastiev,
Symmetry \textbf{9}, 165 (2017).


\bibitem{az11}
A.~V.~Avdeenkov and K.~G.~Zloshchastiev,
J.\ Phys.\ B: At. Mol. Opt. Phys. {\bf 44}, 195303 (2011).



\bibitem{zlo12}
K.~G.~Zloshchastiev, Eur. Phys. J. B \textbf{85}, 273 (2012).


\bibitem{bo15}
B. Bouharia,
Mod. Phys. Lett. B \textbf{29}, 1450260 (2015).

\bibitem{btl16}
V. Bobrov, S. Trigger, and D. Litinski, 
Z. Naturforsch. A \textbf{71}, 565-575 (2016).

\bibitem{z17zna}
K.~G.~Zloshchastiev, 
Z. Naturforsch. A \textbf{72}, 677-687 (2017). 



\bibitem{Zloshchastiev:2009zw}
K.~G.~Zloshchastiev, Grav. Cosmol. \textbf{16}, 288 (2010) [arXiv:0906.4282].


\bibitem{gg15}
K.~G.~Zloshchastiev,
Acta Phys. Polon. B \textbf{42}, 261 (2011) [arXiv:0912.4139].

\bibitem{gg14}
K.~G.~Zloshchastiev,
Phys. Lett. A \textbf{375}, 2305 (2011).


\bibitem{szm16}
T. C. Scott, X. Zhang, R. B. Mann, and G. J. Fee,
Phys. Rev. D \textbf{93}, 084017 (2016).



\bibitem{ch80}
T. Cazenave and A. Haraux, 
Ann.  Fac.  Sci.  Toulouse Math. \textbf{2}, 21–51 (1980).

\bibitem{hh13}
H. Hossieni,
Int. J. Basic Appl. Sci. \textbf{13}, 18 (2013).

\bibitem{gs13}
F. Gladiali and M. Squassina, 
Adv. Nonlinear Stud. \textbf{13}, 663-698 (2013).

\bibitem{dms14}
P. d'Avenia, E. Montefusco, and M. Squassina, 
Commun. Contemp. Math. \textbf{16}, 1350032 (2014).

\bibitem{ss15}
M. Squassina and A. Szulkin,
Calc. Var. \textbf{54}, 585 (2015). 

\bibitem{dsz15}
P. d'Avenia, M. Squassina, and M. Zenari,
Math. Meth. Appl. Sci. \textbf{38}, 5207-5216 (2015).

\bibitem{js16}
C. Ji and A. Szulkin,
J. Math. Anal. Appl. \textbf{437}, 241-254 (2016). 


\bibitem{ard16}
A. H. Ardila, 
Electron. J. Diff. Equat. \textbf{2016: 335}, 1-9 (2016).

\bibitem{ard16a}
A. H. Ardila,
Nonlinear Anal. \textbf{155}, 52-64 (2017).

\bibitem{ard16b}
A. H. Ardila,
Evol. Equ. Control Theory \textbf{6}, 155-175 (2017).

\bibitem{wct16}
W. C. Troy, 
Arch. Rational Mech. Anal. \textbf{222}, 1581-1600 (2016).



\bibitem{tz17}
K. Tanaka and C. Zhang,
Calc. Var. \textbf{56}, 33 (2017).

\bibitem{brz17}
V. Barbu, M. R\"ockner, and D. Zhang,
J. Math. Pures Appl. \textbf{107}, 123-149 (2017).


\bibitem{ns17}
H.-M. Nguyen and  M. Squassina, 
C. R. Acad. Sci. Paris, Ser. I \textbf{355}, 447-451 (2017).

\bibitem{fon17}
F. Fonseca,
Adv. Studies Theor. Phys. \textbf{11}, 105-114 (2017).

\bibitem{pg17}
J. A. Pava and N. Goloshchapova,
Nonlinear Differ. Equ. Appl. \textbf{24}, 27 (2017).


\bibitem{pa17}
J. A. Pava and A. H. Ardila,
Indiana Univ. Math. J., \textbf{67}, 471-494 (2018) [arXiv:1605.05372].




\bibitem{sha17}
L. Shaikhet,
Funct. Differ. Equ. \textbf{24}, 57-67 (2017).

\bibitem{as17}
A. H. Ardila and  M. Squassina, 
Asymptotic Anal. \textbf{107},  203-226 (2018)
[arXiv:1708.03728].

\bibitem{bcs18}
W. Bao, R. Carles, C. Su, and Q. Tang, 
arXiv:1803.10068.


\bibitem{gkz81}
R. G\"ahler, A. G. Klein, and A. Zeilinger, 
Phys. Rev. A \textbf{23}, 1611 (1981).




\bibitem{sha48}
C. E. Shannon,
Bell Syst. Tech. J. \textbf{27}, 
379-423 
(1948).


\bibitem{sha48a}
C. E. Shannon,
Bell Syst. Tech. J. \textbf{27},
623-656 
(1948). 

\bibitem{ever55}
H. Everett III,
``{\it  Theory of the universal wave function},''
PhD thesis, Princeton
(1955) 140 p.


\bibitem{hir57}
I. I. Hirschman, Jr.,
Am. J. Math. \textbf{79}, 152 (1957).



\bibitem{bab61}
K. I. Babenko,
Izv. Akad. Nauk SSSR, Ser. Mat. \textbf{25}, 531 (1961)
[translated in: Amer. Math. Soc. Transl. \textbf{44}, 115 (1961)].

\bibitem{beck75}
W. Beckner,
Annals Math. \textbf{102}, 159-182 (1975).



\bibitem{ps04}
C. J. Pethick and H. Smith,
``\textit{Bose-Einstein Condensation in Dilute Gases},''
Cambridge, UK: CUP (2004) 569p.


\bibitem{gp61} 
E. P. Gross,
Nuov. Cim. \textbf{20}, 454-457 (1961).

\bibitem{gp61a} 
L. P. Pitaevskii,
Sov. Phys. JETP \textbf{13}, 451 (1961).


\bibitem{lan61}
R. Landauer,
IBM J. Res. Dev. \textbf{5}, 183-191 (1961).



\bibitem{bap12}
A. B\'erut, 
A. Arakelyan, A. Petrosyan, S. Ciliberto, R. Dillenschneider, and E. Lutz,
Nature \textbf{483}, 187-190 (2012).

\bibitem{jgb14}
Y. Jun, M. Gavrilov, and J. Bechhoefer,
Phys. Rev. Lett. \textbf{113}, 190601 (2014).

\bibitem{hld16}
J. Hong, B. Lambson, S. Dhuey, and J. Bokor,
Sci. Adv. \textbf{2}, e1501492 (2016). 







\end{thebibliography}
\end{document}